%% file: main.tex
\documentclass[
]{ceurart}

\usepackage[colorinlistoftodos]{todonotes}
\usepackage{listings}
\usepackage{xcolor} 
\usepackage{url}
\usepackage{AMMALanguages}
\usepackage{hyperref}
\usepackage{footnote}
\usepackage{subcaption}
\usepackage{subcaption}
\usepackage{caption}

\usepackage{longtable}
\usepackage{tabularx} 
\usepackage{array} 
\include{commands}
\begin{document}

\copyrightyear{2025}
\copyrightclause{Copyright for this paper by its authors.
  Use permitted under Creative Commons License Attribution 4.0
  International (CC BY 4.0).}

\conference{Joint Proceedings of the STAF 2025 Workshops: OCL, OOPSLE, LLM4SE, ICMM, AgileMDE, AI4DPS, and TTC. Koblenz, Germany, June 10-13, 2025
}

\title{From OCL to JSX: declarative constraint modeling in modern SaaS tools}


\author[]{Antonio Bucchiarone}[%
orcid=0000-0003-1154-1382,
email=antonio.bucchiarone@univaq.it,
url=https://www.disim.univaq.it/AntonioBucchiarone,
]
\cormark[1]
\fnmark[1]
\address[1]{SWEN, Università degli Studi dell'Aquila, 67100, L'Aquila, Italy}

\author[]{Juri Di Rocco}[%
orcid=0000-0002-7909-3902,
email=juri.dirocco@univaq.it,
url=https://www.disim.univaq.it/JuriDiRocco,
]
\fnmark[1]

\author[]{Damiano Di Vincenzo}[%
orcid=0009-0004-7717-0574,
email=damiano.divincenzo@studenti.univaq.it,
]
\fnmark[1]

\author[]{Alfonso Pierantonio}[%
orcid=0000-0002-5231-3952,
email=alfonso.pierantonio@univaq.it,
url=https://www.disim.univaq.it/AlfonsoPierantonio,
]
\fnmark[1]

\cortext[1]{Corresponding author.}
\fntext[1]{These authors contributed equally.}
\begin{abstract}
The rise of Node.js in 2010, followed by frameworks like Angular, React, and Vue.js, has accelerated the growth of low-code development platforms. These platforms harness modern UIX paradigms, component-based architectures, and the SaaS model to enable non-experts to build software. The widespread adoption of single-page applications (SPAs), driven by these frameworks, has shaped low-code tools to deliver responsive, client-side experiences. In parallel, many modeling platforms have moved to the cloud, adopting either server-centric architectures (e.g., GSLP) or client-side intelligence via SPA frameworks, anchoring core components in JavaScript or TypeScript. Within this context, \texttt{OCL.js}, a JavaScript-based implementation of the Object Constraint Language, offers a web-aligned approach to model validation, yet faces challenges such as partial standard coverage, limited adoption, and weak integration with modern front-end toolchains.

In this paper, we explore \texttt{JSX}—a declarative, functional subset of JavaScript/TypeScript used in the React ecosystem - as an alternative to constraint expression in SaaS-based modeling environments. Its component-oriented structure supports inductive definitions for syntax, code generation, and querying. Through empirical evaluation, we compare \texttt{JSX}-based constraints with \texttt{OCL.js} across representative modeling scenarios. Results show \texttt{JSX} provides broader expressiveness and better fits front-end-first architectures, indicating a promising path for constraint specification in modern modeling tools.
\end{abstract}

\begin{keywords}
OCL \sep
JSX \sep
Modeling \sep
SaaS \sep
Low-Code \sep
Jjodel \sep
Model Validation \sep
Declarative Constraint Specification
\end{keywords}

\maketitle

\section{Introduction}

Tools play a pivotal role in the practice and pedagogy of Model-Driven Engineering (MDE)~\cite{schmidt2006model,Kienzle2024}. They support the specification, manipulation, and analysis of models, providing the scaffolding upon which abstractions can be formalized and operationalized. However, despite the increasing relevance of MDE in contemporary software engineering, many of its foundational tools are still rooted in outdated technology stacks. In particular, prominent platforms such as JetBrains MPS\footnote{\url{https://www.jetbrains.com/mps/}} and the Eclipse Modeling Framework\footnote{\url{https://projects.eclipse.org/projects/modeling.emf.emf}} (EMF) were first released prior to 2010, while MetaEdit+\footnote{\url{https://www.metacase.com/products.html}}, one of the first tools of its kind, dates back to 1995. These tools often rely on legacy desktop environments, imperative user interfaces, and monolithic architectures that lack alignment with modern user experience (UIX) paradigms. As a result, they impose a considerable cognitive and technical load on users, particularly those without formal software development background~\cite{bucchiarone2020grand}.

In recent years, the emergence of \emph{single-page applications (SPAs)} has reshaped the landscape of software development, especially within \emph{low-code development platforms} \cite{di2022low}. SPAs provide fluid, dynamic interfaces that run entirely in the browser, minimizing page reloads and latency. This shift has enabled highly responsive and accessible platforms, significantly reducing the entry barrier for end users and promoting the participation of domain experts, non-programmers, and citizen developers~\cite{di2022low}. Frameworks such as {React}\footnote{\url{https://reactjs.org}}, {Angular}\footnote{\url{https://angular.io}}, and {Vue.js}\footnote{\url{https://vuejs.org}}, often in combination with {Node.js}\footnote{\url{https://nodejs.org}}, have played a critical role in this transformation, with React gaining particular prominence due to its robust ecosystem, component-based architecture, and strong developer adoption~\cite{banks2017learning}. According to the 2024 Stack Overflow Developer Survey\footnote{\url{https://survey.stackoverflow.co/2024/technology\#most-popular-technologies-webframe-prof}}, React, Angular, and Vue.js are used together by approximately 77. 6\% professional developers, underscoring their dominant position in the modern front-end development landscape.

In response to these trends, several modern modeling platforms—such as \emph{BESSER}~\cite{besser2024} and \emph{Jjodel}~\cite{jjodel2025,di2023jjodel,di2021enhancing}—have been engineered as single-page applications (SPAs), fully embracing the front-end-first paradigm. This design change enables them to provide rich, responsive user interfaces for tasks such as model editing, validation, and visualization, all within the browser. A cornerstone of many React-based SPAs is {JSX}~\cite{jsxdocs}, a hybrid syntax that seamlessly integrates XML-based HTML with declarative and functional JavaScript constructs. JSX empowers developers to dynamically compose user interfaces and interact fluently with the state of the application~\cite{banks2017learning}, making it a natural fit for model-driven environments built on modern web stacks.

In the context of model-driven environments, the state of the application often encapsulates a variety of modeling artifacts, such as metamodels, user-defined models, and integrity constraints. Within this setting, JSX expressions can be repurposed not only for UI composition but also as expressive mechanisms for \textit{querying} and \textit{validating} these model elements. This dual functionality opens promising avenues for specifying constraints directly within the application logic, bridging the gap between front-end technologies and model semantics. This naturally leads to a fundamental question. 
\begin{quote}
    \emph{Can JSX serve as a viable alternative to traditional constraint languages such as the Object Constraint Language (OCL)}?
\end{quote}

Although OCL remains the standard for expressing model-level constraints in MDE~\cite{cabot2012ocl}, its integration into modern Web environments is hindered by limited support. Tools such as \emph{OCL.js}\footnote{\url{https://github.com/SteKoe/ocl.js}} attempt to bridge this gap by providing a JavaScript-based implementation of OCL, but these tools often suffer from partial standard coverage, weak community support, and fragile maintenance models, which rely on individual developers or small research groups. In contrast, JSX is natively supported in the React ecosystem and is already familiar to many software engineers and students, making it an attractive, low-overhead candidate for constraint specification in React-based modeling platforms.

\smallskip
In this article, we investigate the feasibility and expressiveness of JSX as a constraint language in SPA-based modeling tools. We systematically compare JSX-based expressions with OCL.js through a series of representative modeling scenarios. Our goal is to assess not only the functional equivalence of these approaches but also their ergonomics, maintainability, and alignment with contemporary modeling practices~\cite{alfonso2024ocljs}. The experiments have been conducted by assuming that the artifacts are stored in the application state by means of the Jjodel Object Model (\jjom{})\footnote{\url{https://www.jjodel.io/jjodel-object-model-jjom/}} encoding.

\section{Background}

\subsection{OCL in Modeling: Adaptation to SaaS}
OCL is a formal declarative language designed to specify constraints and queries within modeling environments. Originally introduced as part of the Unified Modeling Language (UML) specification, OCL has become a foundational element in MDE, providing the means to define precise, unambiguous semantics that complement visual modeling notation. It enables developers and modelers to specify a wide range of constraints, including invariants, preconditions, postconditions, and derived attributes, thus enforcing the integrity and correctness of models.
Unlike graphical representations, which often lack the accuracy required for formal validation, OCL introduces a textual layer that supports model verification, validation, and transformation across a broad spectrum of modeling frameworks. One of the key strengths of OCL lies in its powerful collection and navigation operations, which allow modelers to traverse associations, access related elements, and perform complex computations on structured model elements. These capabilities make OCL an indispensable tool in the development of correct-by-construction software systems, especially in safety-critical or rigorously specified domains. OCL is widely used in metamodeling environments like Eclipse EMF\footnote{\url{https://projects.eclipse.org/projects/modeling.mdt.ocl}}, UML, and other MOF-based platforms to enhance the expressiveness and precision of metamodels. It allows defining structural and behavioral constraints on metamodel elements, supporting domain-specific languages (DSLs) and model-driven development (MDD) workflows.
In MDE, the OCL is key to specifying transformation rules and operational semantics, notably in frameworks such as ATL~\cite{jouault2008atl}, QVT~\cite{specificationmof}, and Acceleo\footnote{\url{https://eclipse.dev/acceleo/}}. It is also used to query and ensure semantic correctness throughout the development lifecycle. By adding a formal textual layer to graphical models, OCL enables precise validation, enforces domain rules, and supports traceable, verifiable artifacts - contributing to robust and maintainable model-driven systems.


Low-code platforms have reshaped software development by enabling rapid application creation through visual modeling and reusable components. Most adopt the SaaS model~\cite{choudhary2007software}, leveraging native cloud infrastructures for scalability, collaboration, and integration.
This SaaS shift also affects modeling tools, traditionally desktop-based, which are now evolving toward cloud-based solutions that support real-time collaboration, versioning, and validation, while removing installation and maintenance overhead.
However, OCL remains constrained in web contexts due to limited accessibility and poor integration with JavaScript toolchains. Although MDE is central to defining invariants and behavioral rules, its desktop-centric design hinders adoption in modern SaaS environments.

Addressing this gap requires a JavaScript-based OCL implementation that runs natively in the browser and aligns with declarative, reactive, and event-driven web paradigms. Such an approach would support real-time, client-side constraint validation, offering a more seamless, responsive, and user-friendly experience for both low-code and model-driven web applications.

\subsection{OCL.js}
\input{ocljs}


\subsection{Limitations of OCL.js}

Although OCL.js offers a useful set of features and provides a reasonably comprehensive interpretation of the OCL standard, it reveals several critical limitations, particularly when evaluated in the context of adaptive and reflective platforms such as Jjodel. Initially, Jjodel adopted OCL.js to support model querying and constraint checking. However, as the platform evolved, it became evident that OCL.js could not accommodate the dynamic, introspective requirements of our runtime ecosystem. In response, we developed a dedicated library, \jjom{}, which natively supports querying and navigation of the objects models of Jjodel. The following discussion outlines the key limitations of OCL.js that motivated the design and implementation of \jjom{}.

One of the most fundamental limitations of OCL.js is its incomplete alignment with the official OCL specification, as explicitly acknowledged by the maintainers on the project website\footnote{\url{https://ocl.stekoe.de/\#usage}}:
\begin{quote}
    \textit{This library does not claim to be fully compliant with the OMG OCL definition and might have slight differences.}
\end{quote}
The disclaimer signals the potential for partial feature support, non-standard behaviors, and even semantic inconsistencies, particularly when translating formal OCL specifications into JavaScript environments. Such discrepancies introduce uncertainty into constraint evaluations and diminish confidence in the accuracy of model validation.
Another structural limitation of OCL.js lies in its static expression model. Constraints must be defined and parsed ahead of time, making the library poorly suited for dynamic modeling environments like Jjodel, where models evolve at runtime and constraint definitions may need to be generated or modified on-the-fly. The absence of support for dynamic constraint generation reduces flexibility, hampers maintainability, and limits scalability, particularly in iterative development workflows or systems that involve runtime transformation or user-defined modeling constructs.
OCL.js lacks reflective capabilities, which prevents it from adapting to changes in the underlying metamodel, an essential feature for self-adaptive platforms like Jjodel. Without native reflection, developers must rely on manual updates or external tools, making integration error-prone and fragile.
It also struggles with expressiveness and manageability for complex constraints. Deeply nested logic, chained operations, and advanced collection handling become hard to author and debug. The absence of dedicated tooling or intelligent feedback further reduces usability and increases the risk of validation errors in practice.


Beyond technical limitations, OCL.js faces challenges in sustainability and adoption. Although its GitHub repository\footnote{\url{https://github.com/SteKoe/ocl.js}} shows 762 commits since 2016, development has stagnated. Since September 2021, only 41 of the 269 commits were made by human contributors; the rest were automated updates, indicating minimal active maintenance. Community adoption is also low: a search for \texttt{@stekoe/ocl.js} yields just 26 usage references across 7 public repositories, several of which are inactive. This limited traction raises concerns about OCL.js as a sustainable foundation for reflective modeling platforms. These issues—partial standard coverage, limited adaptability, weak tooling, and declining support—motivate the creation of \jjom{}. Unlike OCL.js, \jjom{} supports runtime introspection, dynamic constraint evaluation, and seamless integration with Jjodel’s object model, offering a more robust and modern alternative.

\section{SaaS-Based Modeling: JSX and the Jjodel Object Model (JjOM)}
The emergence of SaaS-based modeling environments has prompted a rethinking of how constraint definition, model navigation, and runtime interaction are handled in low-code browser-native platforms. In Jjodel, the use of React, Redux\footnote{\url{https://redux.js.org}} and JSX is not limited to the presentation layer; it extends to the fundamental mechanics of modeling and constraint specification. At the heart of this architecture lies the Jjodel Object Model (JjOM), a runtime structure that integrates model data, visualization, and interaction through a modern, declarative web technology stack.

\subsection{React and JSX in Jjodel}
React serves as the backbone of the Jjodel front-end, while JSX, a declarative XML-like syntax embedded in JavaScript/TypeScript\footnote{The JSX version that uses TypeScript is usually called TSX.}, acts as both a UI definition language and a mechanism to interact with model data. In this setting, JSX can be leveraged to express modeling constructs, navigation logic, and even runtime constraints in a style that is readable, composable, and tightly integrated with modern development workflows.
JSX components in Jjodel are inductive building blocks that represent both visual syntax and semantic structure. Each component corresponds to a metamodel concept or model element, and its composition mirrors the hierarchical structure of the modeling languages. These components can be used not only to render visual representations, but also to perform validation and evaluation tasks, enabling declarative logic to be embedded directly into the modeling interface.

Similarly to OCL, the JSX expressions in Jjodel are based on a core set of declarative connectives inherited from JavaScript (see the first table in \cite{jsxconnectives}). These expressions operate in the structured application state, which encodes both models and metamodels according to the internal schema of the \jjom{} as described in the next section. 
Every time the JSX changes, the view is transpiled at run-time into a parametrized component, having all contextual variables (like data) as parameters.
This ensures it is not required to transpile the component again if any variable changes, but it is sufficient to call it as a function to update the GUI.
For further optimization, the view designer can define a list of dependencies that will be displayed in the GUI, and have the component update only if one of those has changed.

\subsection{The Jjodel Object Model (JjOM)}
The \jjom{} provides a structured API and a runtime infrastructure to manage modeling artifacts. It encapsulates both metamodels and models and divides the data into three interconnected submodels.
\begin{itemize}
    \item \textit{Data Submodel}: Represents the abstract structure, including classes (DClass), attributes (DAttribute), references (DReference), objects (DObject) and values (DValue). This submodel forms the core of the Jjodel meta-metamodel and is directly manipulable through the JjOM API.
    \item \textit{Node Submodel}: Captures layout information, including spatial positioning and relationships between visual elements. It also stores validation feedback and serves as a semantic overlay for layout-driven behavior.
    \item \textit{View Submodel}: Handles the concrete syntax and graphical representation of model elements, synchronizing them with the underlying data and node submodels. It also manages constraints and policies that govern model and layout updates, as well as events related to executable models.
\end{itemize}

These submodels ensure consistent handling of abstract models, layout metadata, and visual syntax within a unified native representation of the browser. \jjom{} offers a rich JavaScript API for querying, manipulating, and validating modeling artifacts at runtime, as detailed in~\cite{jsxconnectives}. Through this interface, developers can access and modify class structures, instantiate and navigate object references, define and evaluate constraints using JSX expressions, run queries and transformations throughout the model and perform dynamic validation with \code{validateModel()}. 

For example, the API allows retrieving all attributes of a metaclass (\code{class.attributes}), adding new model instances (\code{addObject()}), and performing filtered queries using JSX-based expressions (\code{executeQuery()}).~An extensive (but not exhaustive) presentation of the object model is given in \cite{jsxconnectives}.

\subsection{JSX for Model Navigation}
In Jjodel, JSX is elevated from a user interface markup language to a fully functional constraint specification mechanism. JSX expressions are used to query and manipulate model artifacts, replacing OCL. An interesting aspect of JSX is that its familiar syntax and declarative structure lower the learning curve, especially for developers and students already proficient in JavaScript. For instance, a JSX expression such as:
\begin{center}
   \code{data.\$ownedAttributes.values.map(a => a.name)}
\end{center}
retrieves the names of all attributes of the selected model element. Here, \code{data} serves a role analogous to \code{self} in OCL, acting as a context pointer to query the active model element. 
In this expression, the \$ prefix is an operator used to invoke reflective access methods provided by the Jjodel runtime. It allows dynamic navigation of model elements and attributes by resolving references at run-time through the Redux-based storage. The \code{.values} accessor retrieves the values from a model collection (in case of a single value the operator is \code{.value}). Finally, the \code{map(a => a.name)} construct is the JavaScript array mapping function, which behaves similarly to the collect operation in OCL.
Typical \texttt{JSX} use cases include accessing and manipulating attribute values, navigating relationships, applying filters or transformations to collections of model elements, executing inline validations directly within the model view, and binding logic to UI events using event-condition-action (ECA) rules.

\begin{figure}[t]
    \centering
    \begin{subfigure}[b]{0.7\textwidth}
        \centering
        \includegraphics[width=1\linewidth]{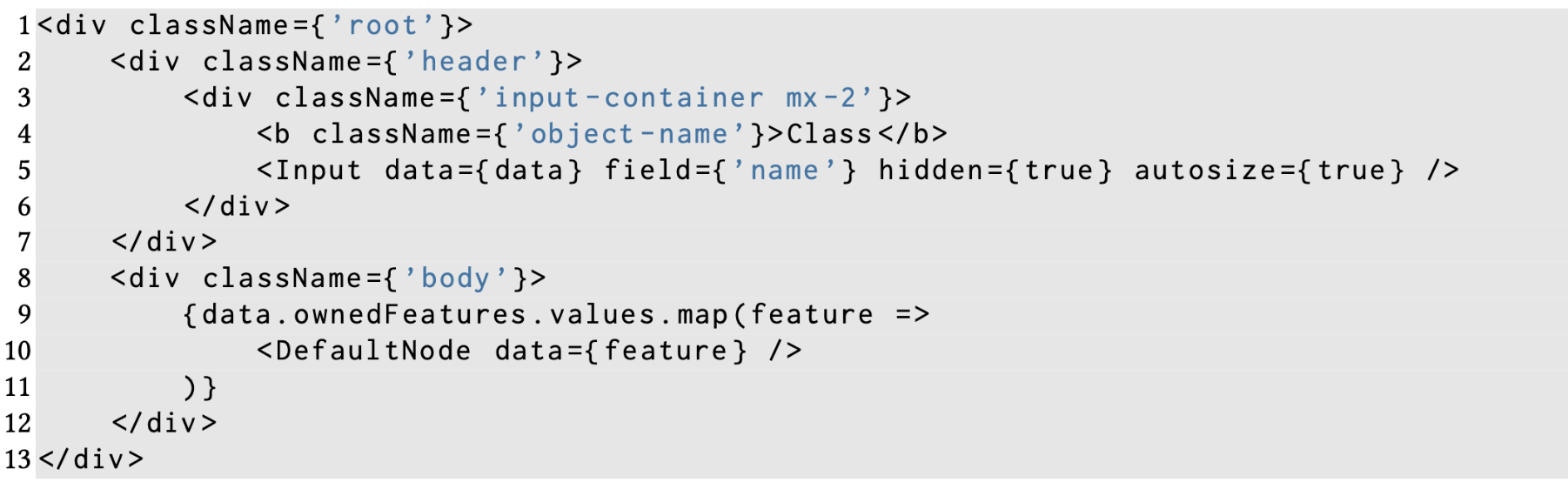}
        \caption{\label{fig:template}The \code{ClassView} template.}
    \end{subfigure}
\quad
    \begin{subfigure}[b]{0.27\textwidth}
        \centering
        \includegraphics[width=1\linewidth]{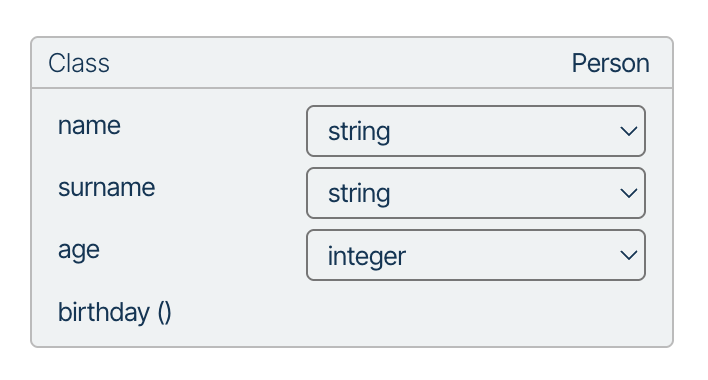}
        \caption{\label{fig:concrete}The \code{Person} class.}
    \end{subfigure}
    \caption{The concrete syntax for \code{Class} elements.}
    \label{fig:concrete-syntax}
\end{figure}

These features enable JSX to seamlessly integrate syntax definition, interaction logic, and constraint validation within a single expressive layer. 
\begin{figure}[t]
   \centering
    \begin{subfigure}[b]{0.45\textwidth}
        \centering
        \includegraphics[width=7.5cm]{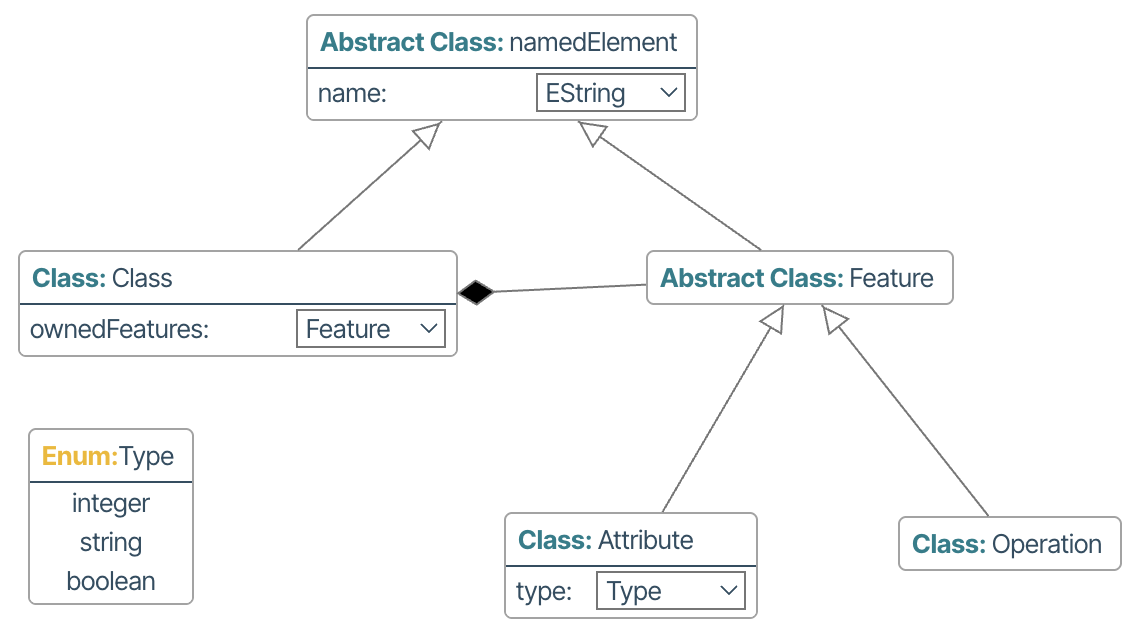}
        \caption{\label{fig:image1}Metamodel.}
        \end{subfigure}
\quad
    \begin{subfigure}[b]{0.45\textwidth}
        \includegraphics[width=7cm]{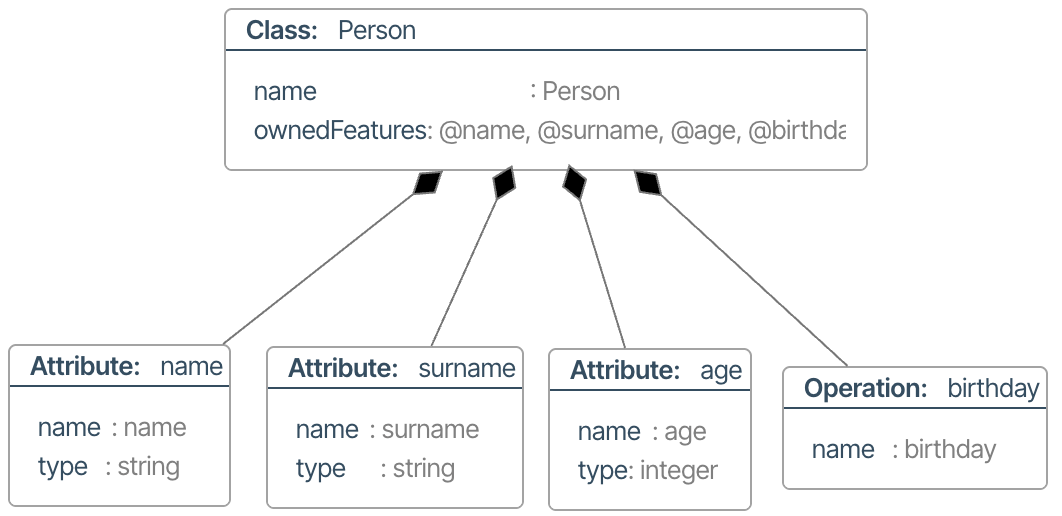}
        \caption{\label{fig:image2}Model.}
    \end{subfigure}
    \caption{\label{fig:UML}UML Class Diagram.}
\end{figure}
To illustrate this capability, consider the metamodel of a simplified UML Class Diagram shown in Figure~\ref{fig:UML}(a), where a \code{Class} is composed of one or more \code{Attribute} and \code{Operation} elements. For the sake of clarity, certain advanced concepts—such as typed parameters for operations and inter-class relationships—have been intentionally omitted. 
An example instance of this metamodel is depicted on the right-hand side of the same figure, where the \code{Person} class contains three attributes: \code{name}, \code{surname}, and \code{age}, together with a single operation \code{birthday}. 
The concrete syntax based on JSX for the element \code{Class}, shown in Figure~\ref{fig:concrete-syntax}(a), defines how instances of the metamodel element \code{Class} are rendered and edited. The component \code{<Input … />} in line 5 enables projectional editing of the class name, allowing users to directly modify the value of the attribute \code{name} through the visual interface. The JSX expression on lines 9-11 dynamically selects all elements of \code{Feature} (i.e. attributes and operations) associated with the class by navigating through \code{data.\$ownedFeatures.values}. These features are then rendered using the \code{<DefaultNode … />} component, which applies the appropriate view to each element based on its type: specifically \code{AttributeView} for attributes and \code{OperationView} for operations\footnote{While this holds in the current example, it represents an oversimplification, views in Jjodel are applied not by type alone, but to instances that satisfy the associated view predicate. This allows for more flexible and context-sensitive rendering beyond simple type-based dispatch}.

This result is obtained by inductively applying the views corresponding to the runtime types of the instances, making JSX highly expressive and adaptable. For instance, to render only attributes and exclude operations, one could use the following JSX expression:

\begin{lstlisting}[caption=Fragment of the OCL.js -Engine and constraints evaluation.,label=lst:jjom-simple]
    {data.$ownedFeatures.values
        .filter(f => f.instanceof.name === 'Attribute')
        .map(a => 
            <DefaultNode data={a} />
        )
    }
\end{lstlisting}

In summary, JSX and the Jjodel Object Model (JjOM) offer a powerful and flexible foundation for defining, navigating, and rendering modeling constructs in modern SaaS-based environments. As illustrated in the previous example by integrating concrete syntax and constraint logic, JSX provides a potential alternative to approaches such as OCL.js. In the following section, we present an empirical evaluation that compares OCL.js and JSX in representative modeling scenarios, assessing their expressiveness, conciseness, and practical usability.

\section{Empirical Validation}
\input{src/evaluation}

\section{Conclusions and Future Works}
The adoption of SPA frameworks like React, Angular, and Vue.js has redefined the architecture of modern software systems, including low-code platforms. In this landscape, modeling tools are shifting toward SaaS-based, browser-native solutions, as exemplified by Jjodel. This evolution demands new approaches to constraint specification beyond traditional, desktop-bound solutions.

\smallskip This paper addressed a central question: \begin{quote} \textit{\textit{Should developers rely on \texttt{OCL.js}, which adapts OCL for the web but lacks full standard coverage and broad support, or adopt \texttt{JSX}, a robust, widely adopted web technology not originally intended for model validation but deeply integrated into the modern web stack?}} \end{quote}

Through an empirical study of \revision{31} modeling scenarios, we found that JSX (via \jjom{}) delivers strong expressiveness and coverage of OCL constructs, with high readability (4.4 vs. 4.0) and effective alignment with modern JavaScript idioms. While some advanced features like \code{@pre} remain unsupported, workarounds exist, and JSX proves to be a promising alternative to \texttt{OCL.js} in SPA-based environments.
Future work will address current gaps by extending \jjom{} with support for state tracking, scoped validation, and a refined API to reduce verbosity while preserving semantic clarity.

\section*{Declaration on Generative AI}

During the preparation of this work, the authors used ChatGPT-4o to grammar and spelling check.

\section*{Acknoledment}
This work has been partially supported by the EMELIOT
national research project, which has been funded by the MUR under the PRIN 2020 program (Contract 2020W3A5FY).

\bibliography{sample-ceur}

\end{document}

%% file: commands.tex
\usepackage{xspace}
\usepackage{tikz}

\newcommand*{\jjodel}{{Jjodel\@\xspace}}
\newcommand*{\jjom}{{JjOM\@\xspace}}
\makeatletter
\newcommand*{\etc}{%
	\@ifnextchar{.}%
	{etc}%
	{etc.\@\xspace}%
}
\makeatother

\newcommand{\code}[1]{\small\texttt{#1}}

\newcommand\revision[1]{\textcolor{blue}{#1}}
\definecolor{codegray}{rgb}{0.5,0.5,0.5}
\definecolor{attributes}{rgb}{0.5,0.5,0.5}
\definecolor{codepurple}{rgb}{0.58,0,0.82}
\definecolor{backcolour}{rgb}{0.95,0.95,0.92}

\definecolor{codered}{HTML}{880044}
\definecolor{codegreen}{HTML}{129490}
\definecolor{codeblue}{HTML}{0A2E36}

\lstdefinestyle{mystyle}{
    backgroundcolor=\color{backcolour},   
    commentstyle=\color{codegreen},
    keywordstyle=\color{codered},
    numberstyle=\tiny\color{codegray},
    stringstyle=\color{codeblue},
    basicstyle=\ttfamily\footnotesize\tiny,
    breakatwhitespace=false,         
    breaklines=true,                 
    captionpos=b,                    
    keepspaces=true,                 
    numbers=left,                    
    numbersep=5pt,                  
    showspaces=false,                
    showstringspaces=false,
    showtabs=false,                  
    tabsize=2,
    keywords={data,value},keywordstyle=\color{codered},
    morekeywords={val,left,right},keywordstyle=\color{codegreen}
}

\definecolor{codered}{HTML}{880044}
\definecolor{codeazure}{HTML}{129490}
\definecolor{codeblue}{HTML}{2660A4}

\lstdefinelanguage{Jjodel}
{
    keywords=[1]{value , DObject, self, instanceof , name},
    keywordstyle=[1]\color{codered},
    keywords=[2]{\$val ,\$left ,\$right },
    keywordstyle=[2]\color{codeazure},
    keywords=[3]{data ,node, view, context, inv},
    keywordstyle=[3]\color{codeblue},
    sensitive=false,
    morestring=[b]',
    morecomment=[l]{--}
}

\usepackage{inconsolata}

\makeatletter
\newcommand{\srcsize}{\@setfontsize{\srcsize}{7pt}{8pt}}
\makeatother
\xdefinecolor{green}{RGB}{63,128,35}
\xdefinecolor{plum}{RGB}{143,0,0}
\xdefinecolor{gray}{RGB}{153,153,153}
\xdefinecolor{red}{RGB}{204,77,101}
\xdefinecolor{blue}{RGB}{51,117,166}
\xdefinecolor{dblue}{RGB}{50,50,200}
\xdefinecolor{commentgreen}{RGB}{30,140,100}

\xdefinecolor{eminence}{RGB}{108,48,130}
\xdefinecolor{weborange}{RGB}{255,165,0}
\xdefinecolor{frenchplum}{RGB}{129,20,83}

\lstdefinelanguage{jsx}{
    basicstyle=\fontfamily{pcr}\srcsize,
    xleftmargin=0pt,
    alsodigit = {-},
    comment=[l]{//},
    morecomment=[s]{/*}{*/},
    commentstyle=\color{commentgreen},
    stringstyle=\color{weborange},
    keywords = {data,node,view},
    keywordstyle=\color{weborange},
    classoffset=1,
    otherkeywords={>,<,.,;,-,!,=,~},
    morekeywords={>,<,.,;,-,!,=,~},
    keywordstyle=\color{plum},
    classoffset=2,
    keywordstyle=\color{plum},
    otherkeywords = {values, oclCondition, name, x, y},
    morekeywords = {values, oclCondition, name, x, y},
    classoffset=3,
    keywordstyle=\color{blue},
    otherkeywords = {ownedAttributes},
    morekeywords = {ownedAttributes},
    classoffset=0
}

\lstdefinelanguage{css}{
    basicstyle=\fontfamily{pcr}\srcsize,
    morecomment=[s]{/*}{*/},
    commentstyle=\color{commentgreen},
    xleftmargin=0pt,
    alsodigit = {-},
    keywords = {grid},
    keywordstyle=\color{green},
    classoffset=1,
    otherkeywords={>,<,.,;,:,!,=,~},
    morekeywords={>,<,.,;,:,!,=,~},
    keywordstyle=\color{gray},
    classoffset=2,
    keywordstyle=\color{plum},
    otherkeywords={15,background-image,background-size,background-position},
    morekeywords={15,background-image,background-size,background-position},
    classoffset=3,
    keywordstyle=\color{red},
    otherkeywords={radial-gradient},
    morekeywords={radial-gradient},
    classoffset=4,
    keywordstyle=\color{plum},
    otherkeywords={0,1,2,3,4,5,6,7,8,9},
    morekeywords={0,1,2,3,4,5,6,7,8,9},
}

\lstdefinelanguage{JavaScript}{
  keywords={break, case, catch, class, const, continue, debugger, default, delete, do, else, export, extends, finally, for, function, if, import, in, instanceof, let, new, return, super, switch, this, throw, try, typeof, var, void, while, with, yield},
  keywordstyle=\color{blue}\bfseries,
  ndkeywords={true, false, null, undefined, NaN, Infinity},
  ndkeywordstyle=\color{teal}\bfseries,
  identifierstyle=\color{black},
  sensitive=true,
  comment=[l]{//},
  morecomment=[s]{/*}{*/},
  commentstyle=\color{gray}\ttfamily,
  stringstyle=\color{red}\ttfamily,
  morestring=[b]',
  morestring=[b]",
}
\lstdefinelanguage{JSX-template}{
    basicstyle=\ttfamily\srcsize,
    comment=[l]{//},
    morecomment=[s]{/*}{*/},
    commentstyle=\color{commentgreen},
    morekeywords = [1]{View, Control, Slider, Input, Selector, Edge, DefaultNode, decorators, values, value, id, node},
    morekeywords = [2]{true, false, div, map},
    morekeywords = [3]{name, ownedAttributes, isPK, left, right},
    morekeywords = [4]{className, field, hidden, autosize, style, start, end, key, data, view, title, payoff, min, max},
    keywordstyle= [1]\color{red},
    keywordstyle = [2]\color{plum},
    keywordstyle = [3]\color{dblue},
    keywordstyle = [4]\color{green},
    sensitive = true,
    morestring = [s]{="}{"},
    morestring = [b]',
    stringstyle = \color{blue},
    literate=%
        {\{data}{\{{\color{red}data\color{black}}}5
        {\$}{{\textcolor{dblue}{\$}}}1%
        {>}{{\textcolor{plum}{>}}}1%
        {=>}{{\textcolor{plum}{=>}}}2%
        {>=}{{\textcolor{plum}{>=}}}2%
        {===}{{\textcolor{plum}{===}}}3%
        {!==}{{\textcolor{plum}{!==}}}3%
        {\&\&}{{\textcolor{plum}{\&\&}}}2%
        {<}{{\textcolor{plum}{<}}}1%
        {/>}{{\textcolor{plum}{/>}}}2%
        {</}{{\textcolor{plum}{</}}}2%
        {=\{}{{\textcolor{plum}{=\{}}}2%
        {\}}{{\textcolor{plum}{\}}}}1%
        {\{}{{\textcolor{plum}{\{\kern-0pt}}}1%
        {)\}}{{\textcolor{plum}{)\}}}}1%
        {)}{{\textcolor{plum}{)}}}1%
        {(}{{\textcolor{plum}{(}}}1%
        {\{data}{{\textcolor{plum}{\{\textcolor{red}{data}}}}5%
        {=\{data}{{\textcolor{plum}{=\{\textcolor{red}{data}}}}5%
}

%% file: ocljs.tex

 OCL.js\footnote{\url{https://ocl.stekoe.de/}} is a JavaScript implementation of OCL for dynamic constraint validation on standard JavaScript objects. It allows a declarative definition of constraints, promoting separation between business logic and validation. As shown in Listing~\ref{lst:evaluationEngine}, after initializing the engine (lines 1–2), constraints are defined on model elements: line 3 ensures that an order contains at least one item; line 4 checks that the total price matches the sum of item prices; line 5 enforces that each item has a positive price. In line 6, an empty order is evaluated, returning \texttt{false} due to constraint violations. Thus, OCL.js supports robust, declarative rule validation within modern web-based modeling environments.

\begin{lstlisting}[breaklines,style=AMMA,language=ATL,mathescape,rulesepcolor=\color{black},captionpos=t,caption=Fragment of the OCL.js -Engine and constraints evaluation.,label=lst:evaluationEngine]
const { OclEngine } = require('ocl.js');
const oclEngine = new OclEngine();
oclEngine.addOclExpression(`context Order inv: self.items->size() > 0`);
oclEngine.addOclExpression(`context Order inv: self.totalPrice = self.items->collect(i | i.price)->sum()`);
oclEngine.addOclExpression(`context Item inv: self.price > 0`);
const order = { id: "O0001", totalPrice : 0, items:[] };
console.log(oclEngine.evaluate(order)); // Returns false (all the constraints are unsadisfied)
\end{lstlisting}

In addition to supporting standard invariants, OCL.js also enables the definition of preconditions and postconditions, which are critical for maintaining transactional integrity in software systems. For instance, in the context of a banking application, a withdrawal operation should only proceed if the account balance remains non-negative after the transaction is applied (see Listing~\ref{lst:ocl-pre-post}). By enforcing such pre- and postconditions, OCL.js helps ensure that critical domain constraints are upheld during function execution. This not only prevents unintended state transitions but also enhances the reliability and correctness of operations in model-driven and data-intensive applications.

\begin{lstlisting}[breaklines,style=AMMA,language=ATL,mathescape,rulesepcolor=\color{black},captionpos=t,caption=Fragment of the OCL.js - Pre and postconditions.,label=lst:ocl-pre-post]
oclEngine.addOclExpression(`
    context Account::withdraw(amount: Number): Boolean
    pre: self.balance >= amount
    post: self.balance = self.balance@pre - amount
`);
\end{lstlisting}

Additionally, OCL.js supports the definition of derived attributes, allowing computed properties to be automatically inferred from existing model data. For example, in a payroll system, an employee’s net salary can be derived from their gross salary and applicable tax deductions, ensuring consistency and reducing redundancy in the model (see Listing~\ref{lst:derived-attribute}). This approach promotes consistency and eliminates redundant computations, thereby reducing the risk of errors in critical domains such as financial calculations. 

\begin{lstlisting}[breaklines,style=AMMA,language=ATL,mathescape,rulesepcolor=\color{black},captionpos=t,caption=Fragment of the OCL.js - Derived attributes.,label=lst:derived-attribute]
oclEngine.addOclExpression(`
    context Employee::netSalary: Number derive: self.grossSalary - self.tax
`);
\end{lstlisting}



Designed to be fully JavaScript native, OCL.js integrates into both front-end and back-end environments. In web applications, it validates client-side forms to ensure user input adheres to business rules, while on the server, it enforces data integrity before database persistence—reducing reliance on database constraints. OCL.js is ideal for graph-based modeling tools and UML editors, enabling real-time validation in domain-specific modeling. Its compatibility with modern JavaScript frameworks makes it well suited for enterprise applications, transactional systems, and model-driven platforms.

By offering a declarative, expressive, and modular approach to constraint specification, OCL.js contributes to building more reliable, verifiable, and maintainable software systems.

%% file: src/evaluation.tex
We compare our proposed model navigation and validation library with OCL.js, using representative OCL statements drawn from literature and practice to illustrate equivalent functionality in \jjom{} and OCL.js. Section~\ref{sec:eval-methodology} outlines the evaluation methodology and comparison criteria. Section~\ref{sec:eval-example} applies these criteria to selected examples, while Section~\ref{sec:eval-reserach-questions} introduces the research questions addressed in Section~\ref{sec:eval-result}..

\subsection{Evaluation methodology and criteria}\label{sec:eval-methodology}

The evaluation methodology includes three steps:

\textit{Dataset Selection}: We selected OCL expressions from "The Ultimate OCL Tutorial"\footnote{\url{https://modeling-languages.com/ocl-tutorial/}}, which lists 56 operations across strings, numerics, booleans, collections, iterators, and \texttt{OclAny}. After reviewing and filtering for clarity and completeness, we retained 31 representative OCL samples. Out of the original 56 operations, we selected 31 by removing duplicates and discarding examples with syntactically incorrect code that did not conform to the grammar. This filtering step ensured the validity and uniqueness of the evaluated operations.

\textit{Implementation of Expressions}: Two co-authors, both proficient in JavaScript and OCL, independently implemented the samples—one using \jjom{}, the other using \texttt{OCL.js}. Due to the lack of metamodels and models, we limited the comparison to \textit{syntax correctness}.

\textit{Measurement and Analysis}: Implementations were evaluated using formal criteria. Metrics were averaged across datasets, and for \texttt{OCL.js}, only the OCL code was considered (excluding API invocation).

\noindent
    $\triangleright$ \textbf{Conciseness} is evaluated by measuring the number of characters (excluding whitespace) (\textbf{COC}). In this comparison, we aim to evaluate \jjom{} in relation to OCL syntax. We are interested in understanding how the combined use of \jjom{} with JSX technology can completely eliminate the need for OCL.js within \jjodel{}. For this reason, we will focus on comparing the effectiveness and conciseness of \jjom{} code against OCL code, without taking into account the API calls from the OCL.js framework. 

\noindent
   $\triangleright$ \textbf{Readability and Understandability} are assessed qualitatively through an expert-based readability assessment (ERA) \cite{oliveira2020evaluating}. We employed a 5-point Likert scale, where 1 indicates poor readability and 5 indicates excellent readability. Each expression was independently evaluated by two experts with practical experience in MDE and proficiency in both OCL and \jjom{}. To ensure consistency and mitigate individual bias, the experts initially rated each expression separately. In cases where discrepancies emerged, the evaluators engaged in a focused discussion to reach a consensus, after which a single agreed-upon score was assigned. This process ensures both independent judgment and mutual calibration, following best practices in qualitative code assessment \cite{oliveira2020evaluating}.

The curated dataset, \jjodel{}, and OCL code, along with the analyzed evaluation criteria, are reported in a Google Sheet.\footnote{\url{https://docs.google.com/spreadsheets/d/1qtgYhGsg9x90H13YdAw-6gk02zP8typD/edit?usp=sharing\&ouid=111608713792271140113\&rtpof=true&sd=true}}



\subsection{Explanatory Example: Equivalent Expressions and Metrics} \label{sec:eval-example}
In this section, we illustrate how the selected evaluation criteria have been applied to the representative OCL.js statements shown in Listings~\ref{lst:evaluationEngine}, \ref{lst:ocl-pre-post}, and \ref{lst:derived-attribute}. In particular, Listing~\ref{lst:jjom-simple} presents the JSX/\jjom{} expressions that realize the same semantics as the OCL.js example shown in Listing~\ref{lst:evaluationEngine}. In \jjodel{}, validation is performed through validation views, where each condition is associated with a dedicated view. When a specified condition is met, the corresponding view is activated, either by displaying an error message or by modifying the graphical representation of the invalid element. The final expression in the list creates an instance of the \code{Order} class, although the same operation can be carried out alternatively through the graphical user interface. Upon creation, the validation view for \code{Item} is activated, rendering its associated feedback based on the violated constraint.

\begin{lstlisting}[caption=Fragment of the OCL.js -Engine and constraints evaluation.,label=lst:jjom-simple]
data.$items.values.length > 0; 
data.$totalPrice.value === data.$items.values.reduce((i, a) => a+=i.$price.value, 0);
data.$price.value > 0
data.parent.addObject({$id: "O0001", $totalPrice: 0, $items: []}, 'Order');
\end{lstlisting}

Replicating Listing~\ref{lst:ocl-pre-post} in \jjodel{} is challenging, as neither \jjodel{} nor \jjom{} natively supports pre- and post-conditions like OCL or \texttt{OCL.js}. While approximations are possible via validation views or event-driven logic, they lack explicit support for \code{@pre}. Nonetheless, \jjodel{}’s architecture could support such features in future extensions through event history or scoped validation.

Listing~\ref{lst:jjom-derived-attribute} shows the \jjom{} version of Listing~\ref{lst:derived-attribute}, using \code{onDataUpdate} and \code{UsageDeclarations} to compute derived values—e.g., \code{netSalary} from \code{grossSalary} and \code{tax}—only when relevant properties change. This selective dependency tracking reduces unnecessary recomputation and improves efficiency, especially in large or dynamic models, offering a precise and maintainable reactive approach.

\begin{lstlisting}[captionpos=t,caption=Fragment of the OCL.js -Engine and constraints evaluation.,label=lst:jjom-derived-attribute]
data.$netSalary.value = data.$grossSalary - data.$tax.value; //placed inside a onDataUpdate event of a view.
\end{lstlisting}


\subsection{Research Questions}\label{sec:eval-reserach-questions}
\noindent
    $\triangleright$ \textbf{RQ$_1$ (Coverage)}: \textit{To what extent does the proposed \jjom{} library support expressing standard OCL statements commonly used in MDE, as represented by examples from "The Ultimate OCL Tutorial"?}
    
    This question directly addresses the completeness and expressiveness of the \jjom{} approach compared to OCL, ensuring that it adequately covers typical modeling scenarios found in practical contexts.

\noindent
    $\triangleright$ \textbf{RQ$_2$ (Effectiveness and Usability):} \textit{How does \jjom{} compare with OCL.js regarding conciseness, structural complexity, readability, and adaptability to metamodel changes when implementing representative model constraints?}

This question specifically targets the metrics outlined in your evaluation criteria, providing a comprehensive comparison of \jjom{} and OCL.js in terms of practical use and maintainability.

\subsection{Results}\label{sec:eval-result}
This section presents the findings from our comparison between \jjom{} and OCL.js, based on the representative dataset derived from The Ultimate OCL Tutorial (see Section~\ref{sec:eval-methodology}). Our analysis addresses the two research questions defined in Section~\ref{sec:eval-reserach-questions} and is guided by the evaluation criteria previously outlined.

\textbf{RQ$_1$}: Our evaluation confirms that \jjom{} is capable of expressing the majority of structural OCL constructs represented in the selected dataset. The analysis was conducted on 31 OCL examples extracted and curated from the OCL Tutorial, covering a broad range of core operations, including invariants, derived attributes, and iterator-based collection expressions.
As a result, our evaluation of \textbf{Expressiveness} is scoped to structural and declarative OCL constructs, which \jjom{} is designed to support directly. Behavioral contracts fall outside the current capabilities of \jjom{} and were not evaluated in this study.
Among the 31 examples, \jjom{} was able to fully express 29 (93.5\%). We were not able to rewrite two OCL statements because they are using per- and post-conditions. Conversely, OCL.js fails in 6 cases (80.6 of examples), primarily due to its lack of native support for global quantification via \code{allInstances} and limited facilities for date manipulation, both of which are present in the dataset.
In conclusion, \jjom{} demonstrates broad coverage and practical expressiveness across standard OCL constructs relevant for structural model validation, with only marginal gaps attributable to its current design focus and the lack of behavioral constraint constructs in the dataset.


\textbf{RQ$_2$}: To assess it, we computed metrics across the implemented expressions in both technologies, focusing on conciseness, and qualitative readability using the representative dataset of 31 OCL expressions described in Section~\ref{sec:eval-methodology}.
Regarding \textbf{Conciseness}, contrary to what might be expected from a JavaScript-based API, \jjom{} expressions were, on average, more concise than their OCL.js equivalents. The mean character count for \jjom{} was 64.8, compared to 74.2 for OCL.js (we calculated the cost of ownership (COC) based on the expression addressed by both approaches). This result suggests that \jjom{}’s idiomatic expression style—despite involving explicit \code{.value} and \code{.values} accessor—still enables more compact constraint definitions. In contrast, OCL.js expressions tend to include verbose constructs (e.g., context declarations, full path navigation, and wrapper methods) that contribute to code bloat, especially in complex model traversals.

\textbf{Readability} was assessed by two independent experts using a 5-point Likert scale. After individual scoring and post-review discussion to resolve discrepancies, an agreed score was recorded for each expression. The average score for OCL.js was 4.4, while \jjom{} scored 4.8, indicating a clear preference for \jjom{} in terms of perceived clarity and developer friendliness. Experts particularly appreciated \jjom{}’s use of familiar JavaScript idioms and its integration within the reactive view-based architecture of \jjodel{}. Although OCL.js offers direct mapping to OCL syntax, it was perceived as less accessible and more syntactically rigid.

These results suggest that \jjom{} offers superior practical usability compared to OCL.js, achieving wider coverage, more concise expressions, and higher readability, all while maintaining comparable structural complexity. For scenarios focused on structural validation in reactive modeling environments, \jjom{} presents itself as a more modern and developer-friendly alternative.